# Navigating the Maize: Cyclic and conditional computational graphs for molecular simulation

Thomas Löhr[1], Michele Assante[2,3], Michael Dodds[1,4], Lili Cao[1], Mikhail Kabeshov[1], Jon-Paul Janet[1], Marco Klähn[1], Ola Engkvist[1,5]

[1] Molecular AI, Discovery Sciences, R&D, AstraZeneca, 431 50 Gothenburg, Sweden

[2] Innovation Centre in Digital Molecular Technologies, Department of Chemistry, University of Cambridge, Lensfield Rd, Cambridge CB2 1EW, UK

[3] Compound Synthesis & Management, The Discovery Centre, Cambridge Biomedical Campus, 1 Francis Crick Avenue, AstraZeneca, CB2 0AA Cambridge, UK

[4] University of St Andrews, KY16 9AJ St Andrews, UK

[5] Department of Computer Science and Engineering, Chalmers University of Technology, Gothenburg, Sweden

## Abstract

Many computational chemistry and molecular simulation workflows can be expressed as graphs. This abstraction is useful to modularize and potentially reuse existing components, as well as provide parallelization and ease reproducibility. Existing tools represent the computation as a directed acyclic graph (DAG), thus allowing efficient execution by parallelization of concurrent branches. These systems can, however, generally not express cyclic and conditional workflows. We therefore developed Maize, a workflow manager for cyclic and conditional graphs based on the principles of flow-based programming. By running each node of the graph concurrently in separate processes and allowing communication at any time through dedicated inter-node channels, arbitrary graph structures can be executed. We demonstrate the effectiveness of the tool on a dynamic active learning task in computational drug design, involving the use of a small molecule generative model and an associated scoring system, and on a reactivity prediction pipeline using quantum-chemistry and semiempirical approaches.

## Introduction
Clearly defined workflows are essential for reproducibility in computational sciences[1]. They make it easier to reason about processes, and allow modularization, fast experimentation, and easy sharing. A workflow can be modelled as a graph, in which each node represents a step of computation, and each edge represents data being passed between steps. We can additionally consider parameters for each node that determine how the computation is performed. As an example, one can view a data processing pipeline as a simple linear workflow, in which data is first read, then processed with a certain set of parameters, and then saved to a new location. Workflows like this are described as *directed acyclic graphs* (DAGs, Figure 1), because they are unidirectional and do not involve cyclic data flows. This means that data flows in one direction only, and each node is only executed a single time. DAGs are a popular model for workflows because they can represent many typical processing tasks, are easy to parallelize using topological sorting[2], and simple enough to reason about. Many tools exist to execute DAGs, popular ones are *Apache Airflow*[3], *Luigi*[4], and *Dagster*[5]. *Knime*[6] is another popular tool, featuring a simplified flow-based architecture optimized for tabular data. Another recent example designed in particular for linear computational chemistry workflows is our tool *Icolos*[7].

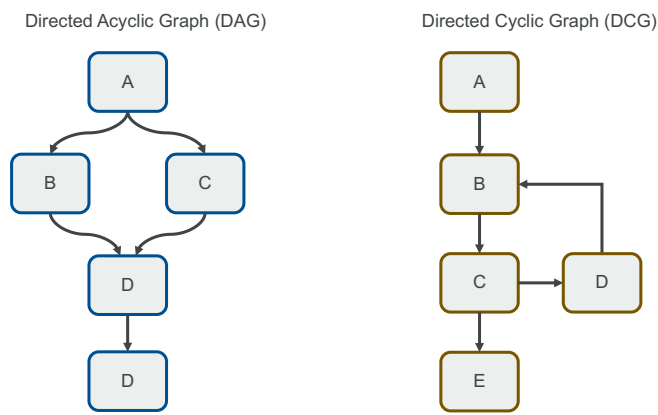

*Figure 1 Directed Acyclic Graphs (DAGs, left) and Directed Cyclic Graphs (DCGs, right). The latter workflow representation allows conditional and iterative execution, common in computational chemistry workflows.*

However, many workflows do not conform to this DAG paradigm, but instead must be modelled as directed (cyclic) graphs (DCGs, Figure 1). This is the case whenever data is passed through the same node repeatedly (without knowing the number of cycles in advance) or passed to different nodes depending on the nature of the data. Because of this, the convenient topological sorting method can no longer be used, so the graph must be modelled differently. One such approach is termed flow-based programming[8,9]. Here, each node in the graph is represented as a separate system process, with data being moved through uni-directional channels. Each node waits for data to be received and can perform computation as soon as all required data has arrived. Thus, every node is essentially independent from and agnostic to the surrounding graph structure. This model of computation has multiple advantages: First, due to each node operating in isolation, unexpected interactions and bugs resulting from different graph structures can be minimized. Second, parallelism is intrinsic to the graph, as each node operates as an independent process and can perform computation as soon as data is available. Third, the use of specific channels as edges makes it easier to reason about data inputs and outputs and provides modularity of components. Possible disadvantages are the potential overhead of many system processes running concurrently, the potentially unclear status of the graph execution (as halting of the computation cannot be readily predicted), and the sometimes-high complexity of the created graphs due to additional data manipulation. Interestingly, this programming model shows strong similarities to digital hardware design, specifically to the concurrent paradigms of hardware description languages such as *Verilog* and *VHDL*.

Here, we developed *Maize*, a workflow manager based on the principles of flow-based programming. The flow-based and non-linear nature sets it apart from our predecessor workflow engine *Icolos*[7]. Maize is written in and interfaces through Python, exposing a simple API to allow users to easily define workflows and add custom nodes. Data handling is accomplished with channels enforcing type safety, thus making the input and output requirements of individual nodes clearer and minimizing the potential for errors during execution. In addition, Maize can handle the sending of both small chunks of data in memory, as well as large files on disk while avoiding race conditions. System and workflow configuration are separated, allowing workflows to be transferable between systems. A feature unique to Maize is that multiple nodes can be grouped together into subgraphs, allowing easier reasoning and node reuse, as well as the construction of highly hierarchical workflows that allow multiple levels of granularity in the workflow specification. An important aspect of Maize, compared to a tool such as *Knime*[6], is the use of Python throughout, including the ability to fully control the workflow execution and resource allocation. This makes it easier to quickly integrate custom software for computational scientists and allows seamless large-scale parallelism. To make the integration into production pipelines as straightforward as possible, workflows can also be specified in JSON or other serialization formats for automated deployments. A final focus has been the tight integration with high-performance computing (HPC) environments, e.g., batch submission systems.

We will first discuss the underlying principles of Maize in more detail, discuss some of the useful emergent properties with regards to processing of large amounts of data and parallelism, and finally demonstrate its use on reinforcement learning and dynamic active learning tasks for early-stage small molecule drug discovery and a reactivity prediction task using quantum-chemical and semiempirical methods.

# Design
## Workflow definition

Maize is written in Python using an object-oriented approach. The computational graph is internally represented in a hierarchical manner as a tree, with the root as the full workflow graph, tree-nodes as (optional) subgraphs, and leaf-nodes as individual computation steps. Each leaf-node (henceforth termed 'node' for brevity) can declare one or more input or output ports representing data receivers and senders respectively, as well as parameters that are static for the duration of graph execution (Figure 2A). The workflow is constructed by first initializing a `Workflow` object, followed by adding individual nodes or predefined subgraphs to the workflow, and finally connecting specific inputs and outputs (Figure 2B). This last step creates a `Channel` object that can pass both files and serialized in-memory data between nodes. At this point, the workflow can be transformed into an executable script, with all node parameters exposed on the command line. Alternatively, the workflow can also be specified using a suitable serialization or configuration system such as JSON or YAML.

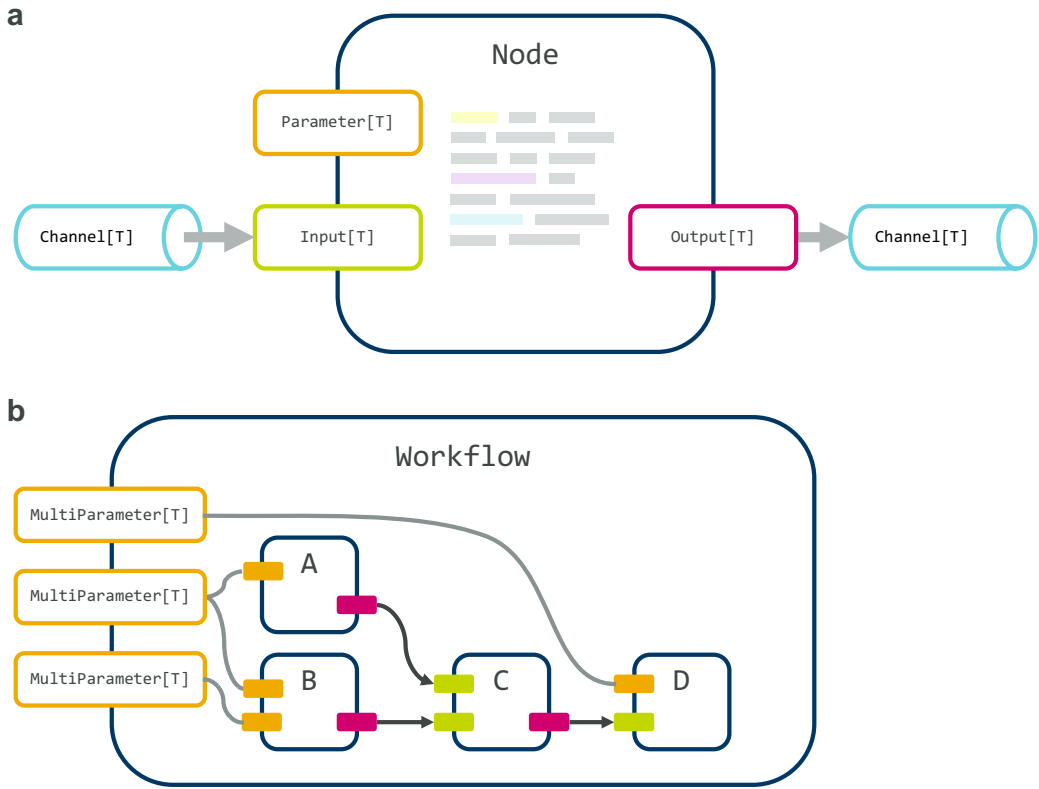

*Figure 2 Node (top) and workflow (bottom) architecture. Nodes expose parameters (static values set prior to execution) and input and / or output ports (allowing data to be passed dynamically). These ports are connected to channels, allowing different nodes to be connected. Workflows can group parameters together and expose them externally, thus abstracting the underlying structure.*

## Execution

Nodes are declared by inheriting from a base Node class, declaring ports and parameters, and defining a `run()` method (Figure 3A). When running the workflow (Figure 3B), each node's `run()` method is executed in a separate process (using Python's `multiprocessing` library), potentially with a different Python interpreter, thus allowing the use of otherwise conflicting environments in a single workflow. Each node will perform any computations it can based on the data available to it through inputs and / or parameters and can send data through its outputs at any time. The `run()` method can either run a single time, causing the node to complete upon returning from the method, or run in a looped mode, i.e., re-running the method upon returning. This latter mechanism allows the creation of cyclic workflows. Conditional execution is possible by sending data to one of multiple outputs, as a result only nodes that receive data will perform computation.

**a**
```python
from maize.core.node import Node
from maize.core.interface import Input, Output
from maize.utilities.chem import IsomerCollection

class Smiles2Molecules(Node):
    """Converts SMILES codes into a set of molecules."""

    inp: Input[list[str]] = Input()
    """SMILES input"""

    out: Output[list[IsomerCollection]] = Output()
    """Molecule output"""

    def run(self) -> None:
        smiles = self.inp.receive()
        mols = [IsomerCollection.from_smiles(smi)
                for smi in smiles]
        self.out.send(mols)
```

**b**
```python
from pathlib import Path
from maize.core.workflow import Workflow
from maize.steps.mai.docking import Vina
from maize.steps.io import LogResult

flow = Workflow(name="docking")
embed = flow.add(Smiles2Molecules)
dock = flow.add(Vina)
result = flow.add(LogResult)

flow.connect_all((embed.out, dock.inp), (dock.out, result.inp))

embed.inp.set(["Nc1nc(F)nc(c12)n(CCCC)c(n2)Cc3cc(OC)ccc3OC"])
dock.receptor.set(Path("./receptor.pdbqt"))
dock.search_center.set((3.3, 11.5, 24.8))
flow.execute()
```

*Figure 3 Maize workflow code. Definition of a custom node embedding a small molecule from a SMILES code (left) and a workflow definition using this node in a linear workflow for docking (right).*

During execution, all nodes communicate their status, log messages, and possible errors to the main parent process through separate message queues. The workflow is stopped if one of the nodes raises an unrecoverable exception, all nodes are completed, or a shutdown signal is set by one of the nodes or an external process. Maize uses several heuristics to determine when to shut down a node, as some nodes may be running in a loop without necessarily performing useful computation. When a node has finished computation and exits, it will close its ports and by extension channels. This closing is communicated to a connected node, which can use its own set of rules to determine if it should also shutdown. Thus, node completion can cascade through the workflow graph.

## Patterns

In the flow-based programming paradigm, some useful patterns can emerge (Figure 4):

- Batch processing: If a very large number of datapoints needs to be processed in a sequential workflow, it can be especially efficient to process it in batches. In Maize, this process is parallel by default, as one batch can be processed on the second node while the next batch is processed on the previous node. An example of this is the process of docking small molecules to a target protein in early-stage drug discovery. The small molecule first needs to be prepared, a process that is typically performed on the CPU, and is then docked, an operation that can often be accomplished on the GPU. Thus, a batch of molecules can be prepared on the CPU, while the previous batch is docking on the GPU.
- Parallelization / load-balancing: Another commonly seen pattern is parallelization. In Maize, this can be accomplished by creating multiple identical workflow branches and distributing the incoming datapoints over all branches. This workflow pattern can be automatically generated and implemented as a subgraph, allowing any kind of computation to be parallelized naturally without having to worry about locks or race conditions.
- Iteration: Many workflows in computational chemistry require performing costly computations until some final condition is fulfilled. This is possible in Maize by creating a node that checks if the computation has completed, sending it either to some final node or back to the computation node for another iteration.

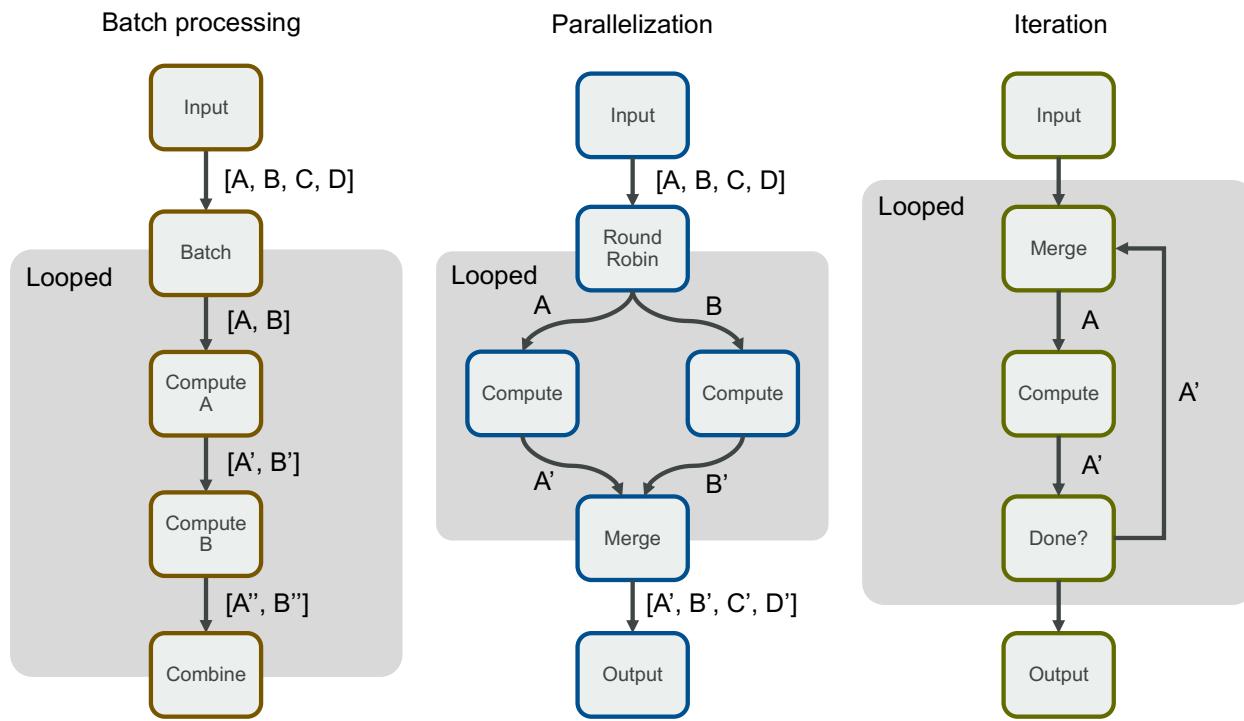

*Figure 4 Useful patterns in flow-based programming. Shaded areas indicate domains of the workflow that are run in a loop, and example data represents the first iteration. Batch processing (left) allows breaking up a large amount of data into chunks, and processing them in parallel, despite the sequential nature of the workflow. Parallelization (middle) allows splitting the data over multiple identical compute nodes. Iteration (right) allows the common pattern of checking a computation for completion and potential re-calculation.*

## Additional features

Maize exposes several convenience functions to make the definition and running of complex workflows easier and more flexible. These include the ability to submit jobs to a queuing system instead of executing locally, re-executing failed nodes multiple times, loading modules (using the LMOD system) from the Python interpreter, automatically connecting nodes based on their port types, renaming and combination of multiple parameters into one, and shortcuts to create for instance the parallelization pattern mentioned above.

## Implemented software

We have implemented interfaces to various software packages common in computational chemistry as Maize nodes. So far, these include quantum chemistry software Gaussian[10], semiempirical packages xTB[11] and CREST[12,13], small molecule docking tools such as, AutoDock-GPU[14], AutoDock Vina[15], GNINA[16], and GLIDE[17], GROMACS[18,19] for molecular dynamics (MD) trajectory analysis, Gypsum-DL[20] for small molecule embedding, and our in-house developed tools REINVENT[21] for AI-based small molecule de-novo drug design and QpTuna, a tool that automatically generates machine learning models for compound property prediction[22], as well as ,various input / output functionality. The domain-agnostic part of Maize also features nodes to enable easier data movement, such as copying, merging, and splitting data. The scope of Maize interfaces is currently expanding rapidly to encompass various tools related to MD simulations including free energy perturbation methods, quantum chemical software and other tools.

# Applications

## *De-novo* design

### Motivation

In this first example we apply Maize on a complex drug discovery workflow, small molecule generation with reinforcement learning[23,24]. The hit-to-lead drug design process typically begins with small molecule hits for a particular target protein. The atomistic structure of these protein-ligand complexes is often available and details the exact position and orientation of the ligand in the protein binding pocket. These initial hit compounds usually exhibit suboptimal properties – they are often not strong and specific binders, and they may have problematic pharmacokinetic properties. It is therefore necessary to find small molecule binders with improved properties using computational approaches, while making use of the information gained from our initial hits. Potential candidates can either be picked from existing compound libraries or created *de novo* using small molecule generative models such as REINVENT[25–27]. The latter method allows guided generation using reinforcement learning[28], i.e., we can feed back a score for each generated molecule indicating if it should be considered favorable or not. As a result, over many iterations, REINVENT will learn to create more suitable molecules. The scoring function used can take many different forms, but here we will be focusing on the docking score, in which a small molecule is fit into a binding pocket by various geometric transformations and the binding energy evaluated using a physics-based approach[29].

### Implementation

We implemented the workflow described above in Maize, using nodes for REINVENT[25], AutoDock GPU[14], Gypsum-DL[20], and various data-handling (Figure 5). The parameter system in Maize allows different configurations of the involved software, as well as changes in how the data is piped through the system. In practice the workflow exhibits some additional complexity: the generated small molecules first need to be embedded, i.e., the SMILES[30] codes need to be converted to an actual 3D representation, which also involves selecting an adequate protonation state and stereo-isomer, for the corresponding compound (using Gypsum-DL[20]). To demonstrate Maize's control flow abilities, we added an additional docking node with higher precision that is triggered whenever the root-mean-square deviation of the docked small molecule to the original reference compound is above a certain threshold.

A flow-based implementation of such a workflow has multiple advantages: First, nodes can be treated completely independently, and are isolated from one another, reducing possible side-effects. Second, the docking node can be re-used in two locations, with the only difference being a slightly different set of parameters. Third, because every node runs in its own process, environments can be kept separate, and code can run in parallel.

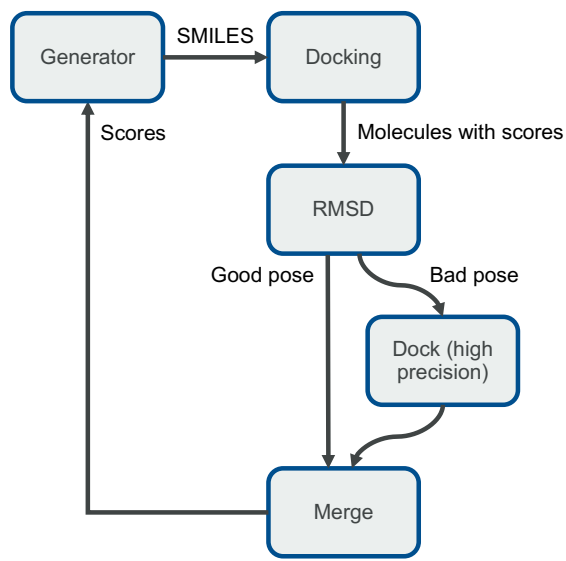

*Figure 5: Small-molecule generation reinforcement learning workflow. Molecules are generated and evaluated by docking them to a target protein structure. Molecules with a large deviation from the reference pose are docked again with higher precision and more conformational sampling. The resulting scores are fed back to the generative model and the process repeated.*

## Active learning

### Motivation

As the number of iterations required to find more favorable small molecules can be quite high, and some scoring methods are often computationally expensive, we would like to replace some of these calculations with a simple machine learning model that can learn an approximation of a physics-based score. This way, instead of always calculating a score using the expensive scoring function, we can in some cases fall back on our fast-to-evaluate approximate model.

This is the main idea behind dynamic active learning[31–33,23,34]: We first generate a set of small molecules to score against our target protein. In the first iteration, these molecules are evaluated using our physics-based *oracle* function such as docking, and the scores fed back to our generator, as well as used to train a simple *surrogate* model emulating our oracle function for future iterations. In subsequent iterations we start by predicting a score for each molecule using this surrogate model. Next, we pick a subset of these compounds using an *acquisition function* to send to our oracle and use the calculated scores to re-train our surrogate model. Finally, we send all scores back to the small molecule generator and repeat the process. This process has large potential savings in computational time, as the accurate but expensive physics-based calculations are reduced. Additionally, the resulting surrogate model can feature high accuracy despite being a simple model such as a random forest due to the very narrow domain. Here, model training will be limited to a single target protein and is thus non-transferable to other targets. Commonly used acquisition functions use various strategies: We could pick a random subset of molecules, pick the ones predicted to have the highest scores (*greedy* sampling), use a combination of both (*epsilon-greedy*), or pick ones with a high uncertainty in their score prediction (e.g., using the *upper-confidence bound*).

### Implementation

Building on the reinforcement learning workflow described above, we implemented an active learning system. We used the same nodes as described above, with the addition of Qptuna[35,36] to provide the surrogate model (Figure 6). The surrogate model is split into separate nodes for training and prediction to simplify the graph dependencies. Finally, the first $n$ iterations will be *pooling* runs to build up the first training dataset for the surrogate model, i.e. during this initial phase all compounds are scored by the oracle only.

As a result of this design, parallelization emerges naturally from the graph definition, for instance, the surrogate model can be re-trained while the next batch of molecules is generated, despite this independence not being explicitly accounted for. The nodes for the tools mentioned above are run in separate Python environments, thus avoiding conflicting dependencies.

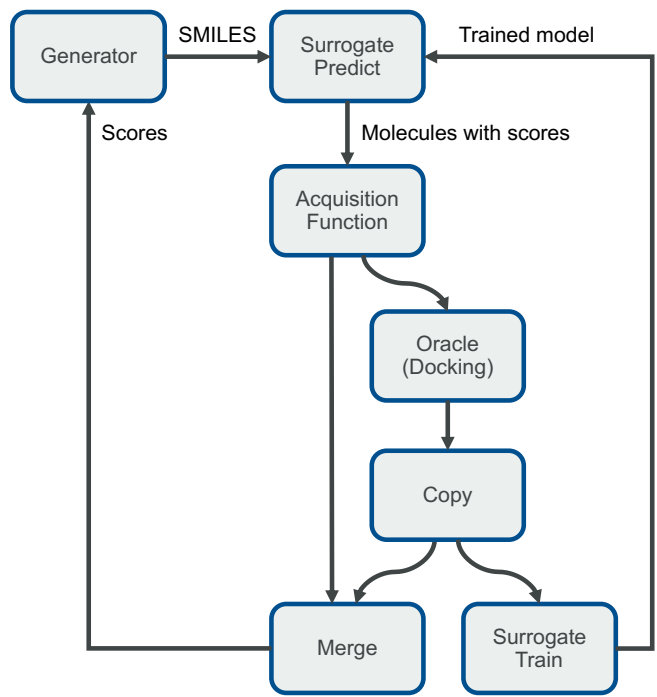

*Figure 6 Simplified active learning workflow. The generator (in this case REINVENT) proposes several molecules, which are fed to a surrogate machine learning model predicting how well these molecules may bind the target protein. Based on these scores, the acquisition function sends a small fraction of the molecules to the computationally expensive oracle (in this case docking). The computed scores are merged with the predicted ones and sent back to the generator to update it. A copy of the scored molecules is sent to the surrogate model for retraining.*

To evaluate the efficiency gains from the above-mentioned parallelism, we ran the active learning workflow in a sequential and fully parallel manner and compared execution times. We limited the run to 10 iterations and used a batch size of 512 generated compounds at each iteration, with 128 acquired molecules to be evaluated by the oracle. The parallel workflow was 13% faster than the naïve sequential workflow due to the more efficient resource utilization. For a more detailed demonstration of the capabilities of dynamic active learning, see ref [34].

## Automated first-principles calculations
### Motivation
First-Principles calculations, such as Quantum-Mechanics and Density Functional Theory, can offer great insight into chemical systems. Molecular properties obtained with such calculations can predict reactivity of chemical compounds and be used as advanced features in data-driven models to increase performances[37–39]. A promising application is the integration of the above-mentioned features in reaction prediction and optimization routines, especially if experimental data is scarce. However, these types of methods often require significant computational resources to be performed as well as specific expertise to be initialized and analyzed correctly. In this context, automation of first-principles calculations could be a remedy to this limitation by providing a faster, more reliable, and more systematic way to perform such calculations. Indeed, to achieve accurate results, a correct description of the system is required; choice of functional, basis set, molecular flexibility and solvent environment are some of the aspects to take in account when defining the system of interest[40]. At the same time, it is crucial to find the right balance between the level of accuracy to be achieved and the computational resources available. In practice, this results in the selection of different computational methods for different tasks, often involving the utilization of multiple software. In this sense, a workflow manager able to orchestrate the requirements of several computational chemistry software is crucial to achieve automation of first-principles calculations and ultimately the integration of such techniques in data-driven methods.

### Implementation

Information about the reaction and the chemical structure of its components is received in tabular format and loaded into a control node. Here, depending on the type of reaction, 3D-geometries for relevant chemical structures in the reaction are generated either through the rdkit package[41] or with custom in-house functions. These input geometries are sent to the first sub-workflow. Initially a molecular mechanics pre-optimisation step removes potential artifacts from the geometries, these are then used as inputs for the conformer generation step performed with semiempirical based metadynamics calculations.

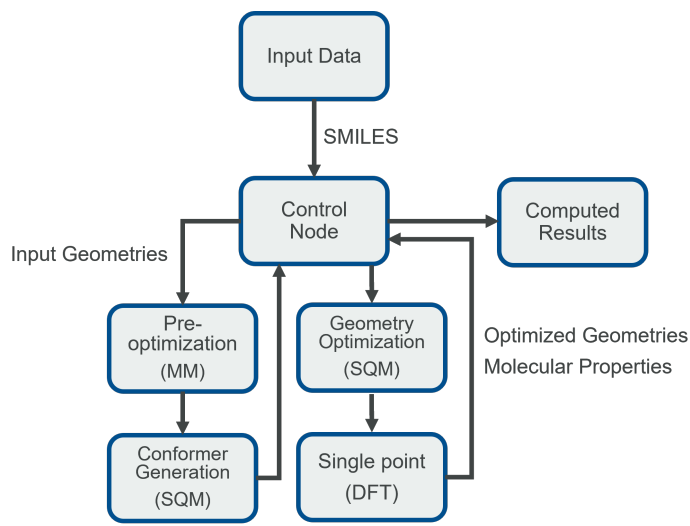

*Figure 7 Reaction prediction workflow based on First-principles calculations. Control node initially generates molecular structures for reactants, products, intermediates, and transition states. Pre-defined structural templates are used to generate geometries for intermediates and templates.*

The generated conformers are later inspected by the control node for any inconsistencies or artifacts and later submitted to the second sub-workflow. Here all the conformers of each component undergo geometry optimization at semiempirical level and single point calculation at DFT level. Once completed, the calculations results are redirected back to the control node which in turn handles potential errors in calculations. Depending on the types of error, individual jobs can be either removed from the workflow or re-submitted. Successful calculations are sent to a return node, which simply reports results in a tabular format.

## Discussion

We have presented Maize, a workflow manager capable of executing cyclic and conditional workflows as commonly found in computational chemistry and early-stage drug discovery. We detailed the design and demonstrated its use on a complex active learning workflow to identify possible new small molecule drug candidates. We showed how parallelization emerges naturally from the graph structure, enabling efficiency improvements in possibly unexpected ways. We also detailed useful patterns providing potentially significant speedups to certain workflows.

While Maize was written with computational chemistry in mind, its architecture and design were deliberately kept domain-agnostic to enable its use in other fields. To enable ease of contributing custom nodes, subgraphs, and workflows, the core domain-agnostic part of Maize is a separate package, and all domain-specific components and utilities are in a separate contribution namespace package. This mechanism allows straightforward extensions and simplifies code reuse.

However, Maize is not necessarily suitable for all workflows: While communication between nodes is fast, it is not intended for low-latency, high-frequency, or inter-processor message passing – here, a system such as the *Message Passing Interface* (MPI) would be more suitable. Related to the previous point is that Maize is not intended to be run on multiple compute nodes the way that MPI applications are, instead Maize can submit jobs to existing job queuing systems such as SLURM and wait for jobs to complete. This means that compute-intensive workflows that potentially require multiple compute nodes will run on a single compute node but submit jobs to other compute nodes and collect the results. Additionally, while we have not observed slowdowns, the use of many Python processes – one for each workflow node – in complex workflows could result in undesirable overheads. The use of threads as an alternative lighter-weight parallelization primitive is currently limited by Python's Global Interpreter Lock (GIL), which prohibits threads from running on multiple processors.

To conclude, we envision Maize as a useful and versatile tool to handle the complexity and many diverse workflows in molecular simulation, computational chemistry, and drug design. It is distributed under the permissive *Apache 2.0* license and available at *https://github.com/MolecularAI/maize* and *https://github.com/MolecularAI/maize-contrib*. The latter includes several prepared workflows for common computational chemistry tasks.

## References

(1) Cohen-Boulakia, S.; Belhajjame, K.; Collin, O.; Chopard, J.; Froidevaux, C.; Gaignard, A.; Hinsen, K.; Larmande, P.; Bras, Y. L.; Lemoine, F.; Mareuil, F.; Ménager, H.; Pradal, C.; Blanchet, C. Scientific Workflows for Computational Reproducibility in the Life Sciences: Status, Challenges and Opportunities. *Future Gener. Comput. Syst.* **2017**, *75*, 284–298. https://doi.org/10.1016/j.future.2017.01.012.

(2) Kahn, A. B. Topological Sorting of Large Networks. *Commun. ACM* **1962**, *5* (11), 558–562. https://doi.org/10.1145/368996.369025.

(3) Apache Airflow, 2023. https://github.com/apache/airflow (accessed 2023-10-17).

(4) Spotify/Luigi, 2023. https://github.com/spotify/luigi (accessed 2023-07-31).

(5) Dagster-Io/Dagster, 2023. https://github.com/dagster-io/dagster (accessed 2023-07-31).

(6) Berthold, M. R.; Cebron, N.; Dill, F.; Gabriel, T. R.; Kötter, T.; Meinl, T.; Ohl, P.; Sieb, C.; Thiel, K.; Wiswedel, B. KNIME: The Konstanz Information Miner. In *Data Analysis, Machine Learning and Applications*; Preisach, C., Burkhardt, H., Schmidt-Thieme, L., Decker, R., Eds.; Studies in Classification, Data Analysis, and Knowledge Organization; Springer Berlin Heidelberg: Berlin, Heidelberg, 2008; pp 319–326. https://doi.org/10.1007/978-3-540-78246-9_38.

(7) Moore, J. H.; Bauer, M. R.; Guo, J.; Patronov, A.; Engkvist, O.; Margreitter, C. Icolos: A Workflow Manager for Structure-Based Post-Processing of de Novo Generated Small Molecules. *Bioinformatics* **2022**, *38* (21), 4951–4952. https://doi.org/10.1093/bioinformatics/btac614.

(8) Morrison, J. P. Data Stream Linkage Mechanism. *IBM Syst. J.* **1978**, *17* (4), 383–408. https://doi.org/10.1147/sj.174.0383.

(9) Morrison, J. P. *Flow-Based Programming, 2nd Edition: A New Approach to Application Development*, 2nd edition.; CreateSpace Independent Publishing Platform: Unionville, Ont., 2010.

(10) Frisch, M. J.; Trucks, G. W.; Schlegel, H. B.; Scuseria, G. E.; Robb, M. A.; Cheeseman, J. R.; Scalmani, G.; Barone, V.; Petersson, G. A.; Nakatsuji, H.; Li, X.; Caricato, M.; Marenich, A. V.; Bloino, J.; Janesko, B. G.; Gomperts, R.; Mennucci, B.; Hratchian, H. P.; Ortiz, J. V.; Izmaylov, A. F.; Sonnenberg, J. L.; Williams; Ding, F.; Lipparini, F.; Egidi, F.; Goings, J.; Peng, B.; Petrone, A.; Henderson, T.; Ranasinghe, D.; Zakrzewski, V. G.; Gao, J.; Rega, N.; Zheng, G.; Liang, W.; Hada, M.; Ehara, M.; Toyota, K.; Fukuda, R.; Hasegawa, J.; Ishida, M.; Nakajima, T.; Honda, Y.; Kitao, O.; Nakai, H.; Vreven, T.; Throssell, K.; Montgomery Jr., J. A.; Peralta, J. E.; Ogliaro, F.; Bearpark, M. J.; Heyd, J. J.; Brothers, E. N.; Kudin, K. N.; Staroverov, V. N.; Keith, T. A.; Kobayashi, R.; Normand, J.; Raghavachari, K.; Rendell, A. P.; Burant, J. C.; Iyengar, S. S.; Tomasi, J.; Cossi, M.; Millam, J. M.; Klene, M.; Adamo, C.; Cammi, R.; Ochterski, J. W.; Martin, R. L.; Morokuma, K.; Farkas, O.; Foresman, J. B.; Fox, D. J. Gaussian 16 Rev. C.01, 2016.

(11) Bannwarth, C.; Caldeweyher, E.; Ehlert, S.; Hansen, A.; Pracht, P.; Seibert, J.; Spicher, S.; Grimme, S. Extended Tight-Binding Quantum Chemistry Methods. *WIREs Comput. Mol. Sci.* **2021**, *11* (2), e1493. https://doi.org/10.1002/wcms.1493.

(12) Pracht, P.; Bohle, F.; Grimme, S. Automated Exploration of the Low-Energy Chemical Space with Fast Quantum Chemical Methods. *Phys. Chem. Chem. Phys.* **2020**, *22* (14), 7169–7192. https://doi.org/10.1039/C9CP06869D.

(13) Grimme, S. Exploration of Chemical Compound, Conformer, and Reaction Space with Meta-Dynamics Simulations Based on Tight-Binding Quantum Chemical Calculations. *J. Chem. Theory Comput.* **2019**, *15* (5), 2847–2862. https://doi.org/10.1021/acs.jctc.9b00143.

(14) Santos-Martins, D.; Solis-Vasquez, L.; Tillack, A. F.; Sanner, M. F.; Koch, A.; Forli, S. Accelerating AutoDock4 with GPUs and Gradient-Based Local Search. *J. Chem. Theory Comput.* **2021**, *17* (2), 1060–1073. https://doi.org/10.1021/acs.jctc.0c01006.

(15) Trott, O.; Olson, A. J. AutoDock Vina: Improving the Speed and Accuracy of Docking with a New Scoring Function, Efficient Optimization, and Multithreading. *J. Comput. Chem.* **2010**, *31* (2), 455–461. https://doi.org/10.1002/jcc.21334.

(16) McNutt, A. T.; Francoeur, P.; Aggarwal, R.; Masuda, T.; Meli, R.; Ragoza, M.; Sunseri, J.; Koes, D. R. GNINA 1.0: Molecular Docking with Deep Learning. *J. Cheminformatics* **2021**, *13* (1), 43. https://doi.org/10.1186/s13321-021-00522-2.

(17) Friesner, R. A.; Banks, J. L.; Murphy, R. B.; Halgren, T. A.; Klicic, J. J.; Mainz, D. T.; Repasky, M. P.; Knoll, E. H.; Shelley, M.; Perry, J. K.; Shaw, D. E.; Francis, P.; Shenkin, P. S. Glide: A New Approach for Rapid, Accurate Docking and Scoring. 1. Method and Assessment of Docking Accuracy. *J. Med. Chem.* **2004**, *47* (7), 1739–1749. https://doi.org/10.1021/jm0306430.

(18) Abraham, M. J.; Murtola, T.; Schulz, R.; Páll, S.; Smith, J. C.; Hess, B.; Lindahl, E. GROMACS: High Performance Molecular Simulations through Multi-Level Parallelism from Laptops to Supercomputers. *SoftwareX* **2015**, *1–2*, 19–25. https://doi.org/10.1016/j.softx.2015.06.001.

(19) Páll, S.; Zhmurov, A.; Bauer, P.; Abraham, M.; Lundborg, M.; Gray, A.; Hess, B.; Lindahl, E. Heterogeneous Parallelization and Acceleration of Molecular Dynamics Simulations in GROMACS. *J. Chem. Phys.* **2020**, *153* (13), 134110. https://doi.org/10.1063/5.0018516.

(20) Ropp, P. J.; Spiegel, J. O.; Walker, J. L.; Green, H.; Morales, G. A.; Milliken, K. A.; Ringe, J. J.; Durrant, J. D. Gypsum-DL: An Open-Source Program for Preparing Small-Molecule Libraries for Structure-Based Virtual Screening. *J. Cheminformatics* **2019**, *11* (1), 34. https://doi.org/10.1186/s13321-019-0358-3.

(21) Loeffler, H. H.; He, J.; Tibo, A.; Janet, J. P.; Voronov, A.; Mervin, L. H.; Engkvist, O. Reinvent 4: Modern AI–Driven Generative Molecule Design. *J. Cheminformatics* **2024**, *16* (1), 20. https://doi.org/10.1186/s13321-024-00812-5.

(22) Mervin, L.; Voronov, A.; Kabeshov, M.; Engkvist, O. QSARtuna: An Automated QSAR Modeling Platform for Molecular Property Prediction in Drug Design. *J. Chem. Inf. Model.* **2024**, *64* (14), 5365–5374. https://doi.org/10.1021/acs.jcim.4c00457.

(23) E. Graff, D.; I. Shakhnovich, E.; W. Coley, C. Accelerating High-Throughput Virtual Screening through Molecular Pool-Based Active Learning. *Chem. Sci.* **2021**, *12* (22), 7866–7881. https://doi.org/10.1039/D0SC06805E.

(24) Filella-Merce, I.; Molina, A.; Orzechowski, M.; Díaz, L.; Zhu, Y. M.; Mor, J. V.; Malo, L.; Yekkirala, A. S.; Ray, S.; Guallar, V. Optimizing Drug Design by Merging Generative AI With Active Learning Frameworks. arXiv May 4, 2023. https://doi.org/10.48550/arXiv.2305.06334.

(25) Blaschke, T.; Arús-Pous, J.; Chen, H.; Margreitter, C.; Tyrchan, C.; Engkvist, O.; Papadopoulos, K.; Patronov, A. REINVENT 2.0: An AI Tool for De Novo Drug Design. *J. Chem. Inf. Model.* **2020**, *60* (12), 5918–5922. https://doi.org/10.1021/acs.jcim.0c00915.

(26) He, J.; Nittinger, E.; Tyrchan, C.; Czechtizky, W.; Patronov, A.; Bjerrum, E. J.; Engkvist, O. Transformer-Based Molecular Optimization beyond Matched Molecular Pairs. *J. Cheminformatics* **2022**, *14* (1), 18. https://doi.org/10.1186/s13321-022-00599-3.

(27) Janet, J. P.; Mervin, L.; Engkvist, O. Artificial Intelligence in Molecular de Novo Design: Integration with Experiment. *Curr. Opin. Struct. Biol.* **2023**, *80*, 102575. https://doi.org/10.1016/j.sbi.2023.102575.

(28) Olivecrona, M.; Blaschke, T.; Engkvist, O.; Chen, H. Molecular De-Novo Design through Deep Reinforcement Learning. *J. Cheminformatics* **2017**, *9* (1), 48. https://doi.org/10.1186/s13321-017-0235-x.

(29) Li, J.; Fu, A.; Zhang, L. An Overview of Scoring Functions Used for Protein-Ligand Interactions in Molecular Docking. *Interdiscip. Sci. Comput. Life Sci.* **2019**, *11* (2), 320–328. https://doi.org/10.1007/s12539-019-00327-w.

(30) Weininger, D. SMILES, a Chemical Language and Information System. 1. Introduction to Methodology and Encoding Rules. *J. Chem. Inf. Comput. Sci.* **1988**, *28* (1), 31–36. https://doi.org/10.1021/ci00057a005.

(31) Sacks, J.; Schiller, S. B.; Welch, W. J. Designs for Computer Experiments. *Technometrics* **1989**, *31* (1), 41–47. https://doi.org/10.1080/00401706.1989.10488474.

(32) Jones, D. R.; Schonlau, M.; Welch, W. J. Efficient Global Optimization of Expensive Black-Box Functions. *J. Glob. Optim.* **1998**, *13* (4), 455–492. https://doi.org/10.1023/A:1008306431147.

(33) Yu, J.; Li, X.; Zheng, M. Current Status of Active Learning for Drug Discovery. *Artif. Intell. Life Sci.* **2021**, *1*, 100023. https://doi.org/10.1016/j.ailsci.2021.100023.

(34) Dodds, M.; Guo, J.; Löhr, T.; Tibo, A.; Engkvist, O.; Paul Janet, J. Sample Efficient Reinforcement Learning with Active Learning for Molecular Design. *Chem. Sci.* **2024**, *15* (11), 4146–4160. https://doi.org/10.1039/D3SC04653B.

(35) Akiba, T.; Sano, S.; Yanase, T.; Ohta, T.; Koyama, M. Optuna: A Next-Generation Hyperparameter Optimization Framework. In *Proceedings of the 25th ACM SIGKDD International Conference on Knowledge Discovery & Data Mining*; KDD '19; Association for Computing Machinery: New York, NY, USA, 2019; pp 2623–2631. https://doi.org/10.1145/3292500.3330701.

(36) QPTUNA: QSAR Using Optimization for Hyper-Parameter Tuning, 2023. https://github.com/MolecularAI/Qptuna (accessed 2023-07-31).

(37) Samha, M. H.; Karas, L. J.; Vogt, D. B.; Odogwu, E. C.; Elward, J.; Crawford, J. M.; Steves, J. E.; Sigman, M. S. Predicting Success in Cu-Catalyzed C–N Coupling Reactions Using Data Science. *Sci. Adv.* **2024**, *10* (3), eadn3478. https://doi.org/10.1126/sciadv.adn3478.

(38) Maley, S. M.; Kwon, D.-H.; Rollins, N.; Stanley, J. C.; Sydora, O. L.; Bischof, S. M.; Ess, D. H. Quantum-Mechanical Transition-State Model Combined with Machine Learning Provides Catalyst Design Features for Selective Cr Olefin Oligomerization. *Chem. Sci.* **2020**, *11* (35), 9665–9674. https://doi.org/10.1039/D0SC03552A.

(39) Jorner, K.; Brinck, T.; Norrby, P.-O.; Buttar, D. Machine Learning Meets Mechanistic Modelling for Accurate Prediction of Experimental Activation Energies. *Chem. Sci.* **2021**, *12* (3), 1163–1175. https://doi.org/10.1039/D0SC04896H.

(40) Bursch, M.; Mewes, J.-M.; Hansen, A.; Grimme, S. Best-Practice DFT Protocols for Basic Molecular Computational Chemistry. *Angew. Chem.* **2022**, *134* (42), e202205735. https://doi.org/10.1002/ange.202205735.

(41) RDKit: Open-Source Cheminformatics. https://www.rdkit.org.